\documentclass{article}%
\usepackage{amssymb}
\usepackage{amsmath}
\usepackage{amsfonts}
\usepackage{graphicx}%
\begin{document}

\title{The Einstein-Hilbert Lagrangian Density in a $2$-dimensional Spacetime is an
Exact Differential\footnote{Published in Modern  Physics Letters {\bf A21} (19), 1519-1527 (2006). }}
\author{{\footnotesize Rold\~{a}o da Rocha}$^{1}${\footnotesize and Waldyr A.
Rodrigues Jr.}$^{2}${\footnotesize \ }\\$^{1}${\footnotesize Instituto de F\'{\i}sica Te\'orica }\\{\footnotesize Universidade Estadual Paulista }\\{\footnotesize Rua Pamplona 145, {}
01405-900 S\~ao Paulo, SP, Brazil}\\{\footnotesize and}\\{\footnotesize DRCC - Institute of Physics Gleb Wataghin }\\{\footnotesize IFGW, UNICAMP CP 6165}\\{\footnotesize 13083-970 Campinas SP, Brasil.  e-mail: {\tt roldao@ifi.unicamp.br}}\\$^{2}${\footnotesize Institute of Mathematics, Statistics and Scientific
Computation }\\{\footnotesize IMECC, UNICAMP, CP 6065}\\{\footnotesize 13083-859 Campinas SP, Brazil. e-mail: {\tt walrod@ime.unicamp.br}}}
\maketitle

\begin{abstract}
Recently Kiriushcheva and Kuzmin \cite{kiku} claimed to have shown that the
Einstein-Hilbert Lagrangian \textit{density} cannot be written in any
coordinate gauge as an exact differential in a $2$-dimensional spacetime.
Since this is contrary to other statements on on the subject found in the
literature, as e.g., by Deser \cite{deser}, Deser and Jackiw
\cite{deserjackiw}, Jackiw \cite{jackiw} and Grumiller, Kummer and Vassilevich
\cite{grumiller} it is necessary to do decide who has reason. This is done in
this paper in a very simply way using the Clifford bundle formalism. In this version we added Section 18 which 
discusses a recent comment on our paper just posted by Kiriushcheva and Kuzmin \cite{kk2}.

\end{abstract}

In \cite{kiku} authors claim to have shown that: `if general covariance is to
be preserved (that is, a coordinate system is not fixed) the well known
triviality of the Einstein field equations in two dimensions is not a
sufficient condition for the Einstein-Hilbert action to be a total
divergence'. This statement is contrary to well known statements, as, e.g., in
\cite{deser,deserjackiw,jackiw,grumiller}). So, we need to decide who is
correct. In what follows we explain that even if at first (and even second)
sight the arguments of \cite{kiku} seems to be correct, they are not complete
and indeed the Einstein-Hilbert Lagrangian in a $2$-dimensional spacetime can
always be written in any coordinate gauge as an exact differential.

To attain our objective in the most efficacious way we shall use in what
follows the Clifford bundle formalism as developed in \cite{rodol2005}, from
where we use the main notations, and where in particular $(M,\mathbf{g}%
,\nabla,\tau_{\mathbf{g}},\uparrow)$ denotes a Lorentzian
spacetime\footnote{$M$ is a $4$-dimensional Hausdorff and paracompact
differentiable manifold, oriented by $\tau_{\mathbf{g}}\in\sec%
{\displaystyle\bigwedge\nolimits^{4}}
T^{\ast}M$ and time oriented by $\uparrow$ (details in \cite{rodol2005}). Also
$\mathbf{g\in}\sec T_{0}^{2}M$ denotes a Lorentz metric of signature $-2$,
$\nabla$ is the Levi-Civita connection of $\mathbf{g}$, and $\mathtt{g}\in\sec
T_{2}^{0}M$ denotes the metric of the cotangent bundle.}. We shall explicitly
calculate the expression of the Einstein-Hilbert Lagrangian density in an
arbitrary chart $(U,\varphi)$ of the maximal atlas of $M$ with coordinate
functions $\{x^{\mu}\}$ using coordinate and orthogonal cobasis and analyze
its behavior in \ a general $n$-dimensional spacetime and in the particular
case of a $2$-dimensional spacetime.

We start by recalling some well known results concerning the differential
geometry associated with a $n$-dimensional Lorentzian manifold, which can be
easily derived using the Clifford bundle of differential forms $\mathcal{C\ell
}(M,\mathtt{g)}$ \cite{rodol2005}\texttt{.}

\textbf{1. }Let $\{e_{\mu}:=\frac{\partial}{\partial x^{\mu}}\}\in\sec
F(M)$\footnote{. Note that each $e_{\mu}\in\sec TM$. i.e., is a vector field.
Also, each $dx^{\mu}\in\sec T^{\ast}M$ and $\mu=1,2,...,n$. In what follows
$F(M)$ denotes the frame bundle and \textbf{P}$_{\mathrm{SO}_{1,n-1}^{e}}(M)$
denotes the orthonormal frame bundle and $P_{\mathrm{SO}_{1,n-1}^{e}}(M)$ the
orthonormal coframe bundle.} a coordinate basis and $\{\vartheta^{\mu}%
=dx^{\mu}\}$ the corresponding dual basis, i.e., $dx^{\mu}(\frac{\partial
}{\partial x^{\nu}})=\delta_{\nu}^{\mu}$. We also have%
\begin{align}
\mathbf{g}(e_{\mu},e_{\nu})  &  =g_{\mu\nu}=g_{\nu\mu}=\mathbf{g}(e_{\nu
},e_{\mu}),\nonumber\\
\mathtt{g}(\vartheta^{\mu},\vartheta^{\nu})  &  =g^{\mu\nu}=g^{\nu\mu
}=\mathtt{g}(\vartheta^{\nu},\vartheta^{\mu}),\label{1}\\
g^{\mu\alpha}g_{\alpha\nu}  &  =\delta_{\nu}^{\mu}.\nonumber
\end{align}

The frame $\{e^{\mu}:=g^{\mu\nu}\frac{\partial}{\partial x^{\nu}}\}\in\sec
F(M)$ is called the reciprocal of the frame $\{e_{\mu}\}$ and the coframe
$\{\vartheta_{\mu}=g_{\mu\nu}\vartheta^{\nu}\}$ is called the
\textit{reciprocal} of the coframe $\{\vartheta^{\mu}\}$, $\vartheta^{\mu}%
\in\sec%
{\displaystyle\bigwedge\nolimits^{1}}
T^{\ast}M\hookrightarrow\sec\mathcal{C\ell}(M,\mathtt{g)}$.

\textbf{2. }We introduce also an orthonormal frame $\{\mathbf{e}_{\mathbf{a}%
}\}\in\sec$\textbf{P}$_{\mathrm{SO}_{1,n-1}^{e}}(M)$ (called a \textit{tetrad
in a }$4$-dimensional spacetime) and \ its dual coframe $\{\theta^{\mathbf{a}%
}\}$ (called a \textit{cotetrad in a }$4$-dimensional spacetime ) which are
basis for $TU$ and $T^{\ast}U$. Using obvious notation we represent the
reciprocal of the frame $\{\mathbf{e}_{\mathbf{a}}\}$ (respectively coframe
$\{\theta^{\mathbf{a}}\}$) by $\{\mathbf{e}^{\mathbf{a}}\}$ (respectively
$\{\theta_{\mathbf{a}}\}$). We \ have\footnote{The boldface indices take the
values \textbf{1}, \textbf{2},...,\textbf{n}.}
\begin{align}
\mathbf{e}_{\mathbf{a}}  &  =h_{\mathbf{a}}^{\mu}e_{\mu}\text{, }%
\mathbf{e}^{\mathbf{a}}=h_{\mu}^{\mathbf{a}}e^{\mu}\text{, }\theta
^{\mathbf{a}}=h_{\mu}^{\mathbf{a}}dx^{\mu}=h_{\mu}^{\mathbf{a}}\vartheta^{\mu
}\text{, }\theta_{\mathbf{a}}=h_{\mathbf{a}}^{\mu}\vartheta_{\mu},\nonumber\\
\mathbf{g}(\mathbf{e}_{\mathbf{a}},\mathbf{e}_{\mathbf{b}})  &  =\eta
_{\mathbf{ab}}=\mathrm{diag}(1,-1,...,-1),\label{2}\\
\mathtt{g}(\theta^{\mathbf{a}},\theta^{\mathbf{b}})  &  :=\theta^{\mathbf{a}%
}\cdot\theta^{\mathbf{b}}=\eta^{\mathbf{ab}}=\mathrm{diag}%
(1,-1,...,-1).\nonumber
\end{align}

\textbf{3. }Define $\theta^{\mathbf{a}_{1}...\mathbf{a}_{r}}=\theta
^{\mathbf{a}_{1}}\wedge...\wedge\theta^{\mathbf{a}_{r}}\in\sec%
{\displaystyle\bigwedge\nolimits^{r}}
T^{\ast}M\hookrightarrow\sec\mathcal{C\ell}(M,\mathtt{g)}$ and ${\star}%
\theta^{\mathbf{a}_{1}...\mathbf{a}_{r}}={\star}(\theta^{\mathbf{a}_{1}}%
\wedge...\wedge\theta^{\mathbf{a}_{r}})\sec%
{\displaystyle\bigwedge\nolimits^{n-r}}
T^{\ast}M\hookrightarrow\sec\mathcal{C\ell}(M,\mathtt{g)}$. In the Clifford
formalism we have for any $A_{r}\in\sec%
{\displaystyle\bigwedge\nolimits^{r}}
T^{\ast}M\hookrightarrow\sec\mathcal{C\ell}(M,\mathtt{g)}$ that%
\begin{equation}
{\star}A_{r}=\tilde{A}_{r}\tau_{\mathbf{g}}=\tilde{A}_{r}\theta^{\mathbf{1}%
...\mathbf{n}}, \label{2bis}%
\end{equation}
where $\tilde{A}_{r}$ denotes the reverse of $A_{r}$, e.g., $\widetilde
{\theta^{\mathbf{a}_{1}}\wedge...\wedge\theta^{\mathbf{a}_{r}}}=\theta
^{\mathbf{a}_{r}}\wedge...\wedge\theta^{\mathbf{a}_{1}}$. \ Observe that we
use the convention that the Clifford product of multiforms is denoted by
juxtaposition of symbols. The following identities which will be used below
are easily shown to be true:%

\begin{align}
d\theta^{\mathbf{a}_{1}...\mathbf{a}_{r}}  &  =-\omega_{\mathbf{b}%
}^{\mathbf{a}_{1}}\wedge\theta^{\mathbf{ba}_{2}...\mathbf{a}_{r}}%
-...-\omega_{\mathbf{b}}^{\mathbf{a}_{r}}\wedge\theta^{\mathbf{a}%
_{1}...\mathbf{a}_{r-1}\mathbf{b}},\label{559n1}\\
d{\star}\theta^{\mathbf{a}_{1}...\mathbf{a}_{r}}  &  =-\omega_{\mathbf{b}%
}^{\mathbf{a}_{1}}\wedge{\star}\theta^{\mathbf{ba}_{2}...\mathbf{a}_{r}%
}-...-\omega_{\mathbf{b}}^{\mathbf{a}_{r}}\wedge{\star}\theta^{\mathbf{a}%
_{1}...\mathbf{a}_{r-1}\mathbf{b}}, \label{559n2}%
\end{align}
where the $\omega_{\mathbf{b}}^{\mathbf{a}}$ are the connection 1-form fields
in a given gauge. We put
\begin{equation}
\omega_{\mathbf{b}}^{\mathbf{a}}:=\omega_{\mathbf{cb}}^{\mathbf{a}}%
\theta^{\mathbf{c}} \label{6}%
\end{equation}

Moreover, we recall that:%

\begin{equation}
\nabla_{\mathbf{e}_{\mathbf{a}}}\mathbf{e}_{\mathbf{b}}:=\omega_{\mathbf{ab}%
}^{\mathbf{c}}\mathbf{e}_{\mathbf{c}},\text{ }\nabla_{\mathbf{e}_{\mathbf{a}}%
}\theta^{\mathbf{b}}:=-\omega_{\mathbf{ac}}^{\mathbf{b}}\theta^{\mathbf{c}}.
\label{7}%
\end{equation}

It is trivial also to show that an analogous formula holds for $\vartheta
^{\mathbf{\alpha}_{1}...\alpha_{r}}=\vartheta^{\alpha_{1}}\wedge
...\wedge\vartheta^{\alpha_{r}}\in\sec%
{\displaystyle\bigwedge\nolimits^{r}}
T^{\ast}M\hookrightarrow\sec\mathcal{C\ell}(M,\mathtt{g)}$ and ${\star
}\vartheta^{\mathbf{\alpha}_{1}...\alpha_{r}}={\star}(\vartheta^{\alpha_{1}%
}\wedge...\wedge\vartheta^{\alpha_{r}})\in\sec%
{\displaystyle\bigwedge\nolimits^{n-r}}
T^{\ast}M\hookrightarrow\sec\mathcal{C\ell}(M,\mathtt{g)}$ with $\omega
_{\mathbf{b}}^{\mathbf{a}_{1}}\mapsto\omega_{\beta}^{\alpha_{1}}=\Gamma
_{\nu\beta}^{\alpha_{1}}\vartheta^{\nu}$, etc., and where $\Gamma_{\nu\beta
}^{\alpha_{1}}=\Gamma_{\beta\nu}^{\alpha_{1}}$ are the Christoffel symbols
defined by:%
\begin{equation}
\nabla_{e_{\mu}}e_{\nu}:=\Gamma_{\mu\nu}^{\alpha}e_{\alpha},\text{ }%
\nabla_{e_{\mu}}\vartheta^{\nu}:=-\Gamma_{\mu\alpha}^{\nu}\vartheta^{\alpha}
\label{n3}%
\end{equation}

\textbf{4. }The following result is a useful one.\textbf{\ }Let ${%
\mbox{\boldmath$\partial$}%
}=d^{\mu}\nabla_{e_{\mu}}=\theta^{\mathbf{a}}\nabla_{\mathbf{e}_{\mathbf{a}}}$
the Dirac operator acting\footnote{The Clifford product is denoted by
juxtaposition of symbols.} on sections of $\mathcal{C\ell}(M,\mathtt{g)}$. If
$A_{p}\in\sec%
{\displaystyle\bigwedge\nolimits^{p}}
T^{\ast}M\hookrightarrow\sec\mathcal{C\ell}(M,\mathtt{g)}$ then%
\begin{align}
{%
\mbox{\boldmath$\partial$}%
}A_{p}  &  ={%
\mbox{\boldmath$\partial$}%
}\mathbf{\wedge}A_{p}+{%
\mbox{\boldmath$\partial$}%
}\mathbf{\cdot}A_{p}\nonumber\\
&  =dA-\delta A_{p}. \label{03}%
\end{align}
In Eq.(\ref{03}) $\delta A_{p}=-{%
\mbox{\boldmath$\partial$}%
}\mathbf{\cdot}A_{p}$ is the Hodge codifferential given by
\begin{equation}
\delta A_{p}=(-)^{p}\star^{-1}d\star A_{p}. \label{003}%
\end{equation}
In particular, if $A\in\sec%
{\displaystyle\bigwedge\nolimits^{1}}
T^{\ast}M\hookrightarrow\sec\mathcal{C\ell}(M,\mathtt{g)}$, writing $A=A_{\mu
}dx^{\mu}=A_{\mathbf{a}}\theta^{\mathbf{a}}$ and \ $\mathbf{a}=\mathtt{g}%
\mathbf{(A,}$ $\mathbf{)}=A^{\mu}e_{\mu}\in\sec TM$ we may verify that the
Hodge codifferential of $A$ is
\begin{align}
\delta A  &  =-{%
\mbox{\boldmath$\partial$}%
}\mathbf{\cdot}A=-\vartheta^{\mu}\cdot\lbrack\nabla_{e_{\mu}}(A_{\nu}%
\vartheta^{\nu})]=-\vartheta^{\mu}\cdot\lbrack(\nabla_{\mu}A_{\nu}%
)\vartheta^{\nu})=-g^{\mu\nu}\nabla_{\mu}A_{\nu}=-(\nabla_{\mu}A^{\mu
})\nonumber\\
&  =-\frac{1}{\sqrt{(-)^{n-1}\det\mathbf{g}}}\partial_{\mu}(\sqrt
{(-)^{n-1}\det\mathbf{g}}A^{\mu}):=-\mathrm{div}\text{ }\mathbf{a} \label{3}%
\end{align}

\textbf{5. }Now, the Einstein-Hilbert Lagrangian density in a $n$-dimensional
Lorentzian spacetime is the $n$-form $\mathcal{L}_{EH}\in\sec%
{\displaystyle\bigwedge\nolimits^{n}}
T^{\ast}M\hookrightarrow\sec\mathcal{C\ell}(M,\mathtt{g)}$%
\begin{align}
\mathcal{L}_{EH}  &  =\frac{1}{2}R\tau_{\mathbf{g}}=R\sqrt{(-1)^{n-1}%
\det\mathbf{g}}dx^{1}\wedge dx^{2}\wedge...\wedge dx^{n}\nonumber\\
&  =\frac{1}{2}R\theta^{\mathbf{1}}\wedge\theta^{\mathbf{2}}\wedge
...\wedge\theta^{\mathbf{n}}=\frac{1}{2}R\theta^{\mathbf{1}}\theta
^{\mathbf{2}}\ldots\theta^{\mathbf{n}}, \label{4}%
\end{align}
where $R$ denotes as usual the curvature scalar. It is a legitimate scalar
function which has, as such, the same value in a given spacetime point when
calculated in any coordinate chart of the maximal of $M.$

\textbf{6.} Cartan's structure equations for a general $n$-dimensional
Lorentzian spacetime are:%
\begin{align}
\mathcal{T}^{\mathbf{a}}  &  =d\theta^{\mathbf{a}}+\omega_{\mathbf{b}%
}^{\mathbf{a}}\wedge\theta^{\mathbf{b}}=0,\nonumber\\
\mathcal{R}_{\mathbf{b}}^{\mathbf{a}}  &  =d\omega_{\mathbf{b}}^{\mathbf{a}%
}+\omega_{\mathbf{c}}^{\mathbf{a}}\wedge\omega_{\mathbf{b}}^{\mathbf{c}},
\label{5}%
\end{align}
where $\omega_{\mathbf{b}}^{\mathbf{a}}$ are the connection 1-form
fields\footnote{Once an orthonormal basis is fixed we suppose for doing
calculations that $\omega_{\mathbf{b}}^{\mathbf{a}}\in\sec%
{\displaystyle\bigwedge\nolimits^{1}}
T^{\ast}M\hookrightarrow\sec\mathcal{C\ell}(M,\mathtt{g).}$} defined by
Eq.(\ref{6}), $\mathcal{T}^{\mathbf{a}}\in\sec%
{\displaystyle\bigwedge\nolimits^{2}}
T^{\ast}M\hookrightarrow\sec\mathcal{C\ell}(M,\mathtt{g)}$ are the torsion
2-form fields and $\mathcal{R}_{\mathbf{b}}^{\mathbf{a}}\in\sec%
{\displaystyle\bigwedge\nolimits^{2}}
T^{\ast}M\hookrightarrow\sec\mathcal{C\ell}(M,\mathtt{g)}$ \ are the curvature
2-form fields. Of course, similar equations can be written using coordinate
basis, in which case the torsion and curvature $2$-forms are denoted by
$\mathcal{T}^{\alpha}(=0)$ and $\mathcal{R}_{\beta}^{\alpha}$.

\textbf{7. }\ With these preliminaries we can rewrite the Einstein-Hilbert
Lagrangian density (in natural units) as:%
\begin{align}
\mathcal{L}_{EH}  &  =\frac{1}{2}R\tau_{\mathbf{g}}\nonumber\\
&  =\frac{1}{2}\mathcal{R}_{\mu\nu}\wedge\star(\vartheta^{\mu}\wedge
\vartheta^{\nu})=\frac{1}{2}\mathcal{R}_{\mathbf{cd}}\wedge\star
(\theta^{\mathbf{c}}\wedge\theta^{\mathbf{d}}) \label{8}%
\end{align}

Indeed, we have immediately using the formulas of Chapter 2 of
\cite{rodol2005}, that
\begin{align}
\mathcal{R}_{\mathbf{cd}}\wedge\star(\theta^{\mathbf{c}}\wedge\theta
^{\mathbf{d}})  &  =(\theta^{\mathbf{c}}\wedge\theta^{\mathbf{d}})\wedge
\star\mathcal{R}_{\mathbf{cd}}=-\theta^{\mathbf{c}}\wedge\star(\theta
^{\mathbf{d}}\lrcorner\mathcal{R}_{\mathbf{cd}})\nonumber\\
&  =-\star\lbrack\theta^{\mathbf{c}}\lrcorner(\theta^{\mathbf{d}}%
\lrcorner\mathcal{R}_{\mathbf{cd}})], \label{8.8}%
\end{align}
and since%
\begin{align}
\theta^{\mathbf{d}}\lrcorner\mathcal{R}_{\mathbf{cd}}  &  =\frac{1}%
{2}R_{\mathbf{cdab}}\theta^{\mathbf{d}}\lrcorner(\theta^{\mathbf{a}}%
\wedge\theta^{\mathbf{b}})=\frac{1}{2}R_{\mathbf{cdab}}(\eta^{\mathbf{da}%
}\theta^{\mathbf{b}}-\eta^{\mathbf{db}}\theta^{\mathbf{a}})\nonumber\\
&  =-R_{\mathbf{ca}}\theta^{\mathbf{b}}=-\mathcal{R}_{\mathbf{c,}} \label{8.9}%
\end{align}
it follows that $-\theta^{\mathbf{c}}\lrcorner(\theta^{\mathbf{d}}%
\lrcorner\mathcal{R}_{\mathbf{cd}})=\theta^{\mathbf{c}}\cdot\mathcal{R}%
_{c}=R.$

\textbf{8. }Now, with a little bit more of algebra we can write the
Einstein-Hilbert Lagrangian density as:%

\begin{align}
\mathcal{L}_{EH}  &  =\mathcal{L}_{g}^{o}-d\left(  \theta^{\mathbf{a}}%
\wedge\star d\theta_{\mathbf{a}}\right) \nonumber\\
&  =\mathcal{L}_{g}^{c}-d\left(  \vartheta^{\alpha}\wedge\star d\vartheta
_{\alpha}\right)  \label{8.14}%
\end{align}
where $\mathcal{L}_{g}^{o}$ and $\mathcal{L}_{g}^{c},$
\begin{equation}
\mathcal{L}_{g}^{o}=-\frac{1}{2}\tau_{\mathbf{g}}\theta^{\mathbf{a}}%
\lrcorner\theta^{\mathbf{b}}\lrcorner\left(  \omega_{\mathbf{ac}}\wedge
\omega_{\mathbf{b}}^{\mathbf{c}}\right)  \text{, \ }\mathcal{L}_{g}^{c}%
=-\frac{1}{2}\tau_{\mathbf{g}}\vartheta^{\alpha}\lrcorner\vartheta^{\beta
}\lrcorner\left(  \omega_{\alpha\gamma}\wedge\omega_{\beta}^{\gamma}\right)  ,
\label{8.15}%
\end{equation}
are first order Lagrangian densities (first introduced by Einstein) written in
\textit{intrinsic} form. Indeed, to prove Eq.(\ref{8.14}) we observe that
using Cartan's second structure equation and Eq.(\ref{559n2}) we can write
$\mathcal{L}_{EH}$ as:%
\begin{align}
\mathcal{L}_{EH}=  &  \frac{1}{2}\star\lbrack\theta^{\mathbf{c}}%
\lrcorner(\theta^{\mathbf{d}}\lrcorner\mathcal{R}_{\mathbf{cd}})]\nonumber\\
&  =\frac{1}{2}\star\lbrack\theta^{\mathbf{c}}\lrcorner(\theta^{\mathbf{d}%
}\lrcorner d\omega_{\mathbf{cd}})]+\text{ }\frac{1}{2}\star\lbrack
\theta^{\mathbf{a}}\lrcorner\theta^{\mathbf{b}}\lrcorner\left(  \omega
_{\mathbf{ac}}\wedge\omega_{\mathbf{b}}^{\mathbf{c}}\right)  ]\nonumber\\
&  =\frac{1}{2}d[\omega_{\mathbf{ab}}\wedge\star(\theta^{\mathbf{a}}%
\wedge\theta^{\mathbf{b}})]+\frac{1}{2}\omega_{\mathbf{ab}}\wedge
d\star(\theta^{\mathbf{a}}\wedge\theta^{\mathbf{b}})+\text{ }\frac{1}{2}%
\star\lbrack\theta^{\mathbf{a}}\lrcorner\theta^{\mathbf{b}}\lrcorner\left(
\omega_{\mathbf{ac}}\wedge\omega_{\mathbf{b}}^{\mathbf{c}}\right)
]\nonumber\\
&  =-d\left(  \theta^{\mathbf{a}}\wedge\star d\theta_{\mathbf{a}}\right)
+\frac{1}{2}\omega_{\mathbf{ab}}\wedge d\star(\theta^{\mathbf{a}}\wedge
\theta^{\mathbf{b}})+\text{ }\frac{1}{2}\star\lbrack\theta^{\mathbf{a}%
}\lrcorner\theta^{\mathbf{b}}\lrcorner\left(  \omega_{\mathbf{ac}}\wedge
\omega_{\mathbf{b}}^{\mathbf{c}}\right)  ]\nonumber\\
&  =-\frac{1}{2}\omega_{\mathbf{ab}}\wedge\omega_{\mathbf{c}}^{\mathbf{a}%
}\wedge\star(\theta^{\mathbf{c}}\wedge\theta^{\mathbf{b}})-d\left(
\theta^{\mathbf{a}}\wedge\star d\theta_{\mathbf{a}}\right)  \label{8.16}%
\end{align}
Also, $\mathcal{L}_{EH}$ can be written as
\begin{equation}
\mathcal{L}_{EH}=\frac{1}{2}\star\lbrack\vartheta^{\gamma}\lrcorner
(\vartheta^{\delta}\lrcorner\mathcal{R}_{\gamma\delta})]=-\frac{1}{2}%
\omega_{\alpha\beta}\wedge\omega_{\gamma}^{\alpha}\wedge\star(\vartheta
^{\gamma}\wedge\vartheta^{\beta})-d\left(  \vartheta^{\mu}\wedge\star
d\vartheta_{\mu}\right)  \label{8.16 new}%
\end{equation}

\bigskip

We now calculate, e.g., $\omega_{\alpha\beta}\wedge\omega_{\rho}^{\alpha
}\wedge\star(\vartheta^{\rho}\wedge\vartheta^{\beta})$. \ We have:%
\begin{align}
\omega_{\alpha\beta}\wedge\omega_{\rho}^{\alpha}\wedge\star(\vartheta^{\rho
}\wedge\vartheta^{\beta})  &  =\vartheta^{\rho}\wedge\vartheta^{\beta}%
\wedge\star(\omega_{\alpha\beta}\wedge\omega_{\rho}^{\alpha})\nonumber\\
&  =\star(\vartheta^{\rho}\wedge\vartheta^{\beta})\cdot(\omega_{\alpha\beta
}\wedge\omega_{\rho}^{\alpha})\nonumber\\
&  \star\lbrack(\vartheta^{\beta}\cdot\omega_{\rho}^{\alpha})(\vartheta^{\rho
}\cdot\omega_{\alpha\beta})-(\vartheta^{\beta}\cdot\omega_{\alpha\beta
})(\vartheta^{\rho}\cdot\omega_{\rho}^{\alpha})]. \label{8.16bis}%
\end{align}
Recalling that $\omega_{\rho}^{\alpha}=\Gamma_{\mu\rho}^{\alpha}\vartheta
^{\mu}$ we get
\begin{equation}
\mathcal{L}_{g}^{c}=-\frac{1}{2}\omega_{\alpha\beta}\wedge\omega_{\gamma
}^{\alpha}\wedge\star(\vartheta^{\gamma}\wedge\vartheta^{\beta})=-\frac{1}%
{2}\tau_{\mathbf{g}}g^{\beta\kappa}\left(  \Gamma_{\kappa\gamma}^{\mu}%
\Gamma_{\mu\beta}^{\gamma}-\Gamma_{\mu\gamma}^{\mu}\Gamma_{\kappa\beta
}^{\gamma}\right)  . \label{8.17bis}%
\end{equation}

Of course, if repeat the calculation using an orthonormal coframe we get
recalling that $\omega_{\mathbf{b}}^{\mathbf{a}}=\omega_{\mathbf{bc}%
}^{\mathbf{a}}\theta^{\mathbf{c}}$ that
\begin{equation}
\mathcal{L}_{g}^{o}=-\frac{1}{2}\omega_{\mathbf{ab}}\wedge\omega_{\mathbf{c}%
}^{\mathbf{a}}\wedge\star(\theta^{\mathbf{c}}\wedge\theta^{\mathbf{b}}%
)=-\frac{1}{2}\tau_{\mathbf{g}}\eta^{\mathbf{bk}}\left(  \omega_{\mathbf{kc}%
}^{\mathbf{d}}\omega_{\mathbf{db}}^{\mathbf{c}}-\omega_{\mathbf{dc}%
}^{\mathbf{d}}\omega_{\mathbf{kb}}^{\mathbf{c}}\right)  . \label{8.17}%
\end{equation}
Eq.(\ref{8.17bis}) shows that $\mathcal{L}_{g}^{o}$ and $\mathcal{L}_{g}^{c}$
are indeed expressions for first order Einstein Lagrangians \textit{density
}in different gauges\textit{. }It is crucial for what follows to realize that
in general \ $\mathcal{L}_{g}^{o}$ $\neq\mathcal{L}_{g}^{c}$. \ Before we use
this fact, let us recall that Eq.(\ref{8.17bis}) shows that, e.g., \ we can
write $\mathcal{L}_{g}^{c}=L_{\Gamma\Gamma}^cd^{n}x$ where
\begin{equation}
L_{\Gamma\Gamma}^{c}=\frac{1}{2}g^{\beta\kappa}\sqrt{(-1)^{n-1}\det\mathbf{g}%
}\left(  \Gamma_{\kappa\gamma}^{\mu}\Gamma_{\mu\beta}^{\gamma}-\Gamma
_{\mu\gamma}^{\mu}\Gamma_{\kappa\beta}^{\gamma}\right)  .
\end{equation}
We notice that as defined $L_{\Gamma\Gamma}^{c}$ is not a \textit{scalar} nor
is it a \textit{scalar density}, to use the wording of old textbooks in
differential geometry and general relativity (see, e.g., \cite{dirac}). There
is a different $L_{\Gamma\Gamma}^{c}$ for every coordinate chart \ that we
choice to use.\ In\cite{kiku} it is claimed that $L_{\Gamma\Gamma}^c$ (or equivalently
$\mathcal{L}_{g}^c$) when expressed in an arbitrary coordinate chart where
$g_{12}\neq0$, is \textit{not} zero, in general, for a 2-dimensional
spacetime, and so that \ the Einstein-Hilbert Lagrangian density cannot be
expressed as an exact differential. Although the statement that in a general
coordinate chart \ $\mathcal{L}_{g}^c\neq 0$ in a $2$-dimensional spacetime is
\textit{correct}, the statement that the Einstein-Hilbert Lagrangian density
cannot be written in a $2$-dimensional spacetime as an exact differential is
\textit{incorrect}.

\textbf{9. }To \ prove our statement, let us first show that\ $\mathcal{L}%
_{g}^o=0$ in a $2$-dimensional spacetime, which implies also that the
corresponding $L_{\omega\omega}^{o}=0$ ($\mathcal{L}_g^o = L_{\omega\omega}^od^2x$) in this case.

Recall that when dim$M=2$ only $\omega_{12}=-\omega_{21}$ is non null. So, the
second member of Eq.(\ref{8.17}) in this case is :%
\begin{align}
-\frac{1}{2}\omega_{\mathbf{ab}}\wedge\omega_{\mathbf{c}}^{\mathbf{a}}%
\wedge\star(\theta^{\mathbf{c}}\wedge\theta^{\mathbf{b}}) &  =-\frac{1}%
{2}\omega_{\mathbf{12}}\wedge\omega_{\mathbf{1}}^{\mathbf{1}}\wedge
\star(\theta^{\mathbf{1}}\wedge\theta^{\mathbf{2}})-\frac{1}{2}\omega
_{\mathbf{12}}\wedge\omega_{\mathbf{2}}^{\mathbf{1}}\wedge\star(\theta
^{\mathbf{2}}\wedge\theta^{\mathbf{2}})\nonumber\\
&  -\frac{1}{2}\omega_{\mathbf{21}}\wedge\omega_{\mathbf{1}}^{\mathbf{2}%
}\wedge\star(\theta^{\mathbf{1}}\wedge\theta^{\mathbf{1}})-\frac{1}{2}%
\omega_{\mathbf{21}}\wedge\omega_{\mathbf{2}}^{\mathbf{2}}\wedge\star
(\theta^{\mathbf{2}}\wedge\theta^{\mathbf{1}})\nonumber\\
&  =0.\label{caceta}%
\end{align}
Next note that although we can write in the general case of a $n$-dimensional spacetime that 
(in obvious notation)
\begin{align}
\mathcal{L}_{EH} &  =-d\left(  \vartheta^{\mu}\wedge\star d\vartheta_{\mu
}\right)  -\frac{1}{2}\omega_{\alpha\beta}\wedge\omega_{\gamma}^{\alpha}%
\wedge\star(\vartheta^{\gamma}\wedge\vartheta^{\beta})\nonumber\\
&  =-d\left(  \theta^{\mathbf{a}}\wedge\star d\theta_{\mathbf{a}}\right)
-\frac{1}{2}\omega_{\mathbf{ab}}\wedge\omega_{\mathbf{c}}^{\mathbf{a}}%
\wedge\star(\theta^{\mathbf{c}}\wedge\theta^{\mathbf{b}}),\label{crucial}%
\end{align}
it is not true, e.g., that $d\left(  \vartheta^{\prime\mu}\wedge\star
d^{\prime}\vartheta_{\mu}\right)  =d\left(  \theta^{\mathbf{a}}\wedge\star
d\theta_{\mathbf{a}}\right)  $ or that $\frac{1}{2}\omega_{\alpha\beta
}^{\prime}\wedge\omega_{\gamma}^{\prime\alpha}\wedge\star(\vartheta
^{\prime\gamma}\wedge\vartheta^{\prime\beta})=\frac{1}{2}\omega_{\mathbf{ab}%
}\wedge\omega_{\mathbf{c}}^{\mathbf{a}}\wedge\star(\theta^{\mathbf{c}}%
\wedge\theta^{\mathbf{b}})$. Only the sums indicated in Eq.(\ref{crucial})
define a \ $n$-form with \textit{tensorial} properties. The parcels are
coordinate gauge dependent as is trivial to verify, and there is no mystery in
this statement, although it may look odd at first sight if these parcels are
written in components.\ Indeed, using Eq.(\ref{2}) we have, e.g.,
\begin{equation}
d\left(  \theta^{\mathbf{a}}\wedge\star d\theta_{\mathbf{a}}\right)  =d\left(
\vartheta^{\mu}\wedge\star d\vartheta_{\mu}\right)  +d\left[  h_{\mu
}^{\mathbf{a}}\vartheta^{\mu}\wedge\star\lbrack(\partial_{\alpha}%
h_{\mathbf{a}}^{\beta})\vartheta^{\alpha}\wedge\vartheta_{\beta}]\right]
\label{crucialbis}%
\end{equation}

\textbf{10.} So, from Eq.(\ref{crucialbis}) and Eq.(\ref{crucial}) we get%
\begin{align}
&  -d\left(  \vartheta^{\mu}\wedge\star d\vartheta_{\mu}\right)  -d\left[
h_{\mu}^{\mathbf{a}}\vartheta^{\mu}\wedge\star\lbrack(\partial_{\alpha
}h_{\mathbf{a}}^{\beta})\vartheta^{\alpha}\wedge\vartheta_{\beta}]\right]
-\frac{1}{2}\omega_{\mathbf{ab}}\wedge\omega_{\mathbf{c}}^{\mathbf{a}}%
\wedge\star(\theta^{\mathbf{c}}\wedge\theta^{\mathbf{b}})\nonumber\\
&  =-d\left(  \vartheta^{\mu}\wedge\star d\vartheta_{\mu}\right)  -\frac{1}%
{2}\omega_{\alpha\beta}\wedge\omega_{\gamma}^{\alpha}\wedge\star
(\vartheta^{\gamma}\wedge\vartheta^{\beta}).\label{crux}%
\end{align}

\textbf{11.} Now, since in a $2$-dimensional spacetime Eq.(\ref{caceta}) says
that $\frac{1}{2}\omega_{\mathbf{ab}}\wedge\omega_{\mathbf{c}}^{\mathbf{a}%
}\wedge\star(\theta^{\mathbf{c}}\wedge\theta^{\mathbf{b}})=0$ we get (in this
case)%
\begin{equation}
d\left[  h_{\mu}^{\mathbf{a}}\vartheta^{\mu}\wedge\star\lbrack(\partial
_{\alpha}h_{\mathbf{a}}^{\beta})\vartheta^{\alpha}\wedge\vartheta_{\beta
}]\right]  =\frac{1}{2}\omega_{\alpha\beta}\wedge\omega_{\gamma}^{\alpha
}\wedge\star(\vartheta^{\gamma}\wedge\vartheta^{\beta}).\label{cruxbis}%
\end{equation}

With this result we can write the Einstein-Hilbert Lagrangian density in a
$2$-dimensional spacetime (denoted $\mathcal{L}_{EH}^{(2)}$) as an exact
differential, i.e., \
\begin{equation}
\mathcal{L}_{EH}^{(2)}=-d\left[  \vartheta^{\mu}\wedge\star d\vartheta_{\mu
}+h_{\mu}^{\mathbf{a}}\vartheta^{\mu}\wedge\star\lbrack(\partial_{\alpha
}h_{\mathbf{a}}^{\beta})\vartheta^{\alpha}\wedge\vartheta_{\beta}]\right]
,\label{fim}%
\end{equation}
which is the result that we wanted to show. We observe that the final
expression (i.e., Eq.( \ref{fim})) needs the introduction of a tetrad field to
be written, but it is true in an arbitrary coordinate chart.

\textbf{12. }We observe also that Eq.(\ref{cruxbis}) shows explicitly the
error done by authors of Ref. \cite{kiku}. Indeed, they affirm that the term
$-\frac{1}{2}\omega_{\alpha\beta}\wedge\omega_{\gamma}^{\alpha}\wedge
\star(\vartheta^{\gamma}\wedge\vartheta^{\beta})=-\frac{1}{2}\tau_{\mathbf{g}%
}g^{\beta\kappa}\left(  \Gamma_{\kappa\gamma}^{\mu}\Gamma_{\mu\beta}^{\gamma
}-\Gamma_{\mu\gamma}^{\mu}\Gamma_{\kappa\beta}^{\gamma}\right)  $ cannot be
written as an exact differential, which as we just saw, is not the case.

\textbf{13. }It is also worth to note that in an orthonormal gauge the form of
the Einstein-Hilbert Lagrangian in a $2$-dimensional spacetime is simply
\begin{equation}
\mathcal{L}_{EH}^{(2)}=\frac{1}{2}R\tau_{\mathbf{g}}=-d\left(  \theta
^{\mathbf{a}}\wedge\star d\theta_{\mathbf{a}}\right)  .\label{EHagain}%
\end{equation}
Then we can write taking into account that in a $2$-dimensional spacetime
$\tau_{\mathbf{g}}^{2}=1$ we have taking into account Eq.(\ref{2bis}) that \
\begin{equation}
\sqrt{-\det\mathbf{g}}R=-2\sqrt{-\det\mathbf{g}}d\left(  \theta^{\mathbf{a}%
}\wedge\star d\theta_{\mathbf{a}}\right)  \tau_{\mathbf{g}}=2\sqrt
{-\det\mathbf{g}}\star d\left(  \theta^{\mathbf{a}}\wedge\star d\theta
_{\mathbf{a}}\right)  ,\label{gru}%
\end{equation}
an equation identical to Eq.(1.55) of \cite{grumiller}, establishing the
correspondence of the formalisms. Also, Eq.(\ref{gru}) shows that the
statement in \cite{deser} that \ first oder theory does not involve the
zweibein $\{\theta^{\mathbf{a}}\}$ is not correct. Indeed, in the formula used
in \cite{deser} the $\{\theta^{\mathbf{a}}\}$ disappeared after using Cartan's
first structure equation in Eq.(\ref{gru}).

\textbf{14. }We now obtain a very convenient form for $\mathcal{L}_{g}^o$ ($\mathcal{L}_g^c$)  in
a $n$-dimensional spacetime in terms of $\{\theta^{\mathbf{a}}\}$ ($\{\vartheta^{\mathbf{a}}\}$), which may
be appropriately called the Thirring Lagrangian
\cite{rodol2005,quin,quinrod,thirring}. Note that for the case, e.g., of $\mathcal{L}_g^o$, we first verify using
Cartan's first structure equation that%
\begin{equation}
\omega^{\mathbf{cd}}=\frac{1}{2}\left[  \theta^{\mathbf{d}}\lrcorner
d\theta^{\mathbf{c}}-\theta^{\mathbf{c}}\lrcorner d\theta^{\mathbf{d}}%
+\theta^{\mathbf{c}}\lrcorner\left(  \theta^{\mathbf{d}}\lrcorner
d\theta_{\mathbf{a}}\right)  \theta^{\mathbf{a}}\right]  .\label{8.19}%
\end{equation}
Using Eq.(\ref{8.19}) in Eq.(\ref{8.15}) we get%
\begin{align}
\mathcal{L}_{g}^o &  =-\frac{1}{2}\tau_{\mathbf{g}}\theta^{\mathbf{a}}%
\lrcorner\theta^{\mathbf{b}}\lrcorner\{\frac{1}{2}[\theta_{\mathbf{a}%
}\lrcorner d\theta_{\mathbf{c}}+\theta_{\mathbf{c}}\lrcorner d\theta
_{\mathbf{a}}+\theta_{\mathbf{a}}\lrcorner(\theta_{\mathbf{c}}\lrcorner
d\theta_{\mathbf{k}})\theta^{\mathbf{k}}]\nonumber\\
&  \wedge\frac{1}{2}[\theta_{\mathbf{b}}\lrcorner d\theta^{\mathbf{c}}%
+\theta^{\mathbf{c}}\lrcorner d\theta_{\mathbf{b}}+\theta^{\mathbf{c}%
}\lrcorner(\theta_{\mathbf{b}}\lrcorner d\theta^{\mathbf{l}})\theta
_{\mathbf{l}}],\label{8.20}%
\end{align}
which after some algebraic manipulations using, e.g., the identities of
Chapter 2 of \cite{rodol2005}, reduces to
\begin{equation}
\mathcal{L}_{g}^o=-\frac{1}{2}(d\theta^{\mathbf{a}}\wedge\theta
^{\mathbf{b}})\wedge\star(d\theta_{\mathbf{b}}\wedge\theta_{\mathbf{a}}%
)+\frac{1}{4}\left(  d\theta^{\mathbf{a}}\wedge\theta_{\mathbf{a}}\right)
\wedge\star\left(  d\theta^{\mathbf{b}}\wedge\theta_{\mathbf{b}}\right)
\label{8.20bis}%
\end{equation}
or equivalently,%
\begin{equation}
\mathcal{L}_{g}^o=-\frac{1}{2}d\theta^{\mathbf{a}}\wedge\star d\theta
_{\mathbf{a}}+\frac{1}{2}\delta\theta^{\mathbf{a}}\wedge\star\delta
\theta_{\mathbf{a}}+\frac{1}{4}\left(  d\theta^{\mathbf{a}}\wedge
\theta_{\mathbf{a}}\right)  \wedge\star\left(  d\theta^{\mathbf{b}}%
\wedge\theta_{\mathbf{b}}\right)  .\label{8.1bis}%
\end{equation}

Eq.(\ref{8.1bis}) is the basis for a possible formulation of gravitational
theory in Minkowski spacetime, once we introduce the concept of deformation
extensor fields. Details are to be found in \cite{rodol2005}

\textbf{15. }The expression for $\mathcal{L}_{g}^o$ given, e.g., by
Eq.(\ref{8.20bis}) (or Eq.(\ref{8.1bis})) obviously defines a unique $n$-form
when expressed in any coordinate chart. Since it does not contain the gauge
dependent connection $1$-forms $\omega_{\mathbf{b}}^{\mathbf{a}}$ and since
moreover it is \textit{obviously }null \footnote{Recall, e.g., that
$d\theta_{\mathbf{b}}\wedge\theta_{\mathbf{a}}$ and $\left(  d\theta
^{\mathbf{a}}\wedge\theta_{\mathbf{a}}\right)  $ in Eq.(\ref{8.20bis}) are
$3$-forms in a $2$-dimensional spacetime and then are null.} for a
$2$-dimensional spacetime someone may may think equivocally (as we did at a
first sight) that the fact that $\mathcal{L}_{EH}$ is indeed an exact
differential in that case, does not imply that it must be an exact
differential when expressed in an arbitrary coordinate gauge.

\textbf{16. }For completeness it remains to calculate the \ `divergence term'
using Clifford algebras methods in a $2$-dimensional spacetime, since such an
exercise shows how powerful and economic is this calculation instrument.

We have,
\begin{align}
\vartheta^{\mathbf{\mu}}\wedge\star d\vartheta_{\mathbf{\mu}}  &
=-\star(\vartheta^{\mathbf{\mu}}\lrcorner d\vartheta_{\mathbf{\mu}%
})\nonumber\\
&  =-\star\lbrack\vartheta^{\mathbf{\mu}}\lrcorner d(g_{\mu\alpha}%
\vartheta^{\alpha})]\nonumber\\
&  =-\star\lbrack\vartheta^{\mathbf{\mu}}\lrcorner(g_{\mu\alpha},_{\beta
}\vartheta^{\beta}\wedge\vartheta^{\alpha})]\nonumber\\
&  =-\star\lbrack g_{\mu\alpha},_{\beta}g^{\mu\beta}\vartheta^{\alpha}%
-g_{\mu\alpha},_{\beta}g^{\mu\alpha}\vartheta^{\beta}]\nonumber\\
&  =-\star\lbrack g^{\mu\beta}\left(  g_{\mu\alpha},_{\beta}-g_{\mu\beta
},_{\alpha}\right)  \vartheta^{\alpha}]\label{11}\\
&  =-\star\lbrack g^{\mu\beta}g^{\nu\alpha}\left(  g_{\mu\alpha},_{\beta
}-g_{\mu\beta},_{\alpha}\right)  \vartheta_{\alpha}]:=-\star\lbrack A^{\mu
}\vartheta_{\mu}]=-\star A.\nonumber
\end{align}
and
\begin{equation}
h_{\mu}^{\mathbf{a}}\vartheta^{\mu}\wedge\star\lbrack(\partial_{\alpha
}h_{\mathbf{a}}^{\beta})\vartheta^{\alpha}\wedge\vartheta_{\beta}=-\star
h_{\alpha}^{\mathbf{a}}(\partial^{\alpha}h_{\mathbf{a}}^{\beta}-\partial
^{\beta}h_{\mathbf{a}}^{\alpha})\vartheta_{\beta}=-\star B, \label{11bis}%
\end{equation}
with $A,B\in\sec%
{\displaystyle\bigwedge\nolimits^{1}}
T^{\ast}M.$

Now,
\begin{align}
-d\left(  \vartheta^{\mathbf{\mu}}\wedge\star d\vartheta_{\mathbf{\mu}%
}\right)   &  =d\star A=(-1)\star(-1)\star^{-1}d\star A=-\star\delta
A,\label{12}\\
-d[h_{\mu}^{\mathbf{a}}\vartheta^{\mu}\wedge\star\lbrack(\partial_{\alpha
}h_{\mathbf{a}}^{\beta})\vartheta^{\alpha}\wedge\vartheta_{\beta}]  &
=-d\star B=-\star\delta B
\end{align}

Finally, \ calling $A+B=V\in%
{\displaystyle\bigwedge\nolimits^{1}}
T^{\ast}M$ we get%

\begin{equation}
-d\left(  \vartheta^{\mathbf{\mu}}\wedge\star d\vartheta_{\mathbf{\mu}}%
+h_{\mu}^{\mathbf{a}}\vartheta^{\mu}\wedge\star\lbrack(\partial_{\alpha
}h_{\mathbf{a}}^{\beta})\vartheta^{\alpha}\wedge\vartheta_{\beta}\right)
=\partial_{\mu}\left(  \sqrt{-\det\mathbf{g}}V^{\mu}\right)  d^{2}x=\frac
{1}{2}R\tau_{g}, \label{16}%
\end{equation}
which furnishes an alternative expression for $R$ relative to Eq.(1.49) of
\cite{grumiller} or Eq.(2.8) of \ \cite{deserjackiw}.

\textbf{17. }We recall that this result implies that unfortunately there is no
consistent Hamiltonian formulation for the Einstein-Hilbert action in
$2$-dimensional spacetime, contrary to what is stated in \cite{kiku} and also
in \cite{kiku1,kiku2}. Indeed, as showed, e.g., in \cite{martinec} the
Einstein-Hilbert action is a topological invariant \ (in Euclidean signature
it is \ directly related to the genus of a $2$-dimensional Riemannian space
\cite{grumiller}). So the spatial metric has no conjugate momentum and the
canonical formalism breaks down. We observe however, that as showed by
Polyakov \ \cite{polyakov} in quantum field theory there are subtle effects
which gives rise to a nontrivial effective action. It is this effective action
that is used in the quantization of $2$-dimensional gravity.

\textbf{18. }Kiriushcheva and Kuzmin have just posted comments on our paper \cite{kk2}. They argued that our recent result,  
that {}\\{}\bigskip
\begin{center}
\begin{minipage}{11cm}``\emph{in two dimensional spacetime the Lagrangian of tetrad gravity is an exact differential [1], 
despite the claim of the authors, neither proves the Jackiw conjecture [2], nor contradicts to the conclusion of [3]
This demonstrates that the tetrad formulation is different from the metric formulation of the Einstein-Hilbert action.}'' \footnote{Of course 
the references [1-3] in the quotation are the ones in \cite{kk2}.} \end{minipage}\end{center}

Our paper shows without doubt that the Einstein-Hilbert Lagrangian density in any dimension can be trivially written in terms of tetrads, their differentials
and the Hodge star operator. Moreover we give a proof that in 2-dimensional spacetime the Einstein-Hilbert Lagrangian
is an exact differential. The criticisms of Kiriushcheva and Kuzmin are all ill founded, but to be fair we admit that there are  misprints
in Sections 8,  9 and 14, where labels $c$ and $o$ are missing on the Lagrangian densities $\mathcal{L}_g^o$ and $\mathcal{L}_g^c$. 

However, contrary to what they claimed we never wrote that ``$\mathcal{L}_{\Gamma\Gamma}^c = L^o_{\omega\omega}$''. Indeed, just the opposite is true
as the reader can verify reading carefully Sections 8 and 9 of previous versions of the present paper, \emph{even without the correction of the misprints}! 

We would like also to comment on footnote 4 of \cite{kk2}, where the authors say that with the Clifford algebra methods we get two different 
results for the problem of the Einstein-Hilbert Lagrangian in a 2-dimensional spacetime in two days.

Well, only one of the results is correct, and it is the one which appears in the second version and of course, also in this version. In the first version we commit a lapse which conduced to a wrong conclusion, as we mention in Section 15 above. To end we mention 
that we sent both versions to Kiriushcheva and Kuzmin, \emph{before} posting them. However these authors did not send their comments on our 
paper to us.

\end{document}